%% file: FinalVersion.tex
\documentclass[journal]{vgtc}         

\ifpdf
  \pdfoutput=1\relax                   
  \usepackage{graphicx}                
  \DeclareGraphicsExtensions{.pdf,.png,.jpg,.jpeg} 
\else
  \usepackage{graphicx}                
  \DeclareGraphicsExtensions{.eps}     
\fi%

\graphicspath{{figures/}{pictures/}{images/}{./}} 

\usepackage{microtype}                 
\PassOptionsToPackage{warn}{textcomp}  
\usepackage{textcomp}                  
\usepackage{mathptmx}                  
\usepackage{times}                     
\usepackage{cite}                      
\usepackage{tabu}                      
\usepackage{booktabs}                  
\usepackage{xspace}
\usepackage{xcolor}
\usepackage{enumitem}
\usepackage{amsmath}
\usepackage{subfiles}
\usepackage{overpic}
\usepackage{ccicons}
\usepackage{float}
\usepackage{stfloats}
\usepackage{textcomp}
\usepackage{tikz}
\usepackage{orcidlink}

\newcommand{\etal}{et~al.\xspace}

\definecolor{myblue}{rgb}{0.1,0.3,0.8}

\onlineid{1227}

\vgtccategory{Research}
\vgtcpapertype{application/design study}

\title{3DStoryline: Immersive Visual Storytelling}

\author{{Haonan Yao}\orcidlink{0009-0003-1735-6281}, {Lixiang Zhao}\orcidlink{0000-0001-6181-1673}, Boyuan Chen\orcidlink{0009-0006-9645-4526}, Kaiwen Li\orcidlink{0009-0003-2064-6992}, {Hai-Ning Liang}\orcidlink{0000-0003-3600-8955}, {Lingyun Yu}\orcidlink{0000-0002-3152-2587}}

\authorfooter{
\item Haonan Yao, Lixiang Zhao, Boyuan Chen, Kaiwen Li and Lingyun Yu are with Department of Computing, Xi'an Jiaotong-Liverpool University, Suzhou, China.
\item Hai-Ning Liang is with Computational Media and Arts Thrust, the Hong Kong University of Science and Technology (Guangzhou), Guangzhou, China.
 \item The first two authors contribute equally to this work.
 \item Lingyun Yu is the corresponding author. Email: Lingyun.Yu@xjtlu.edu.cn.
}

\abstract{
Storyline visualization has emerged as an innovative method for illustrating the development and changes in stories across various domains. Traditional approaches typically represent stories with one line per character, progressing from left to right. While effective for simpler narratives, this method faces significant challenges when dealing with complex stories involving multiple characters, as well as temporal and spatial dynamics. In this study, we investigate the potential of immersive environments for enhancing storyline visualizations. We begin by summarizing the key design considerations for effective storyline visualization in virtual reality (VR). Guided by these principles, we develop 3DStoryline, a system that allows users to view and interact with 3D immersive storyline visualizations. To evaluate the effectiveness of 3DStoryline, we conduct a task-based user study, revealing that the system significantly enhances users' comprehension of complex narratives.
} 


\CCScatlist{ 
 \CCScat{K.6.1}{Management of Computing and Information Systems}%
{Project and People Management}{Life Cycle};
 \CCScat{K.7.m}{The Computing Profession}{Miscellaneous}{Ethics}
}

\teaser{
  \centering
  \begin{tikzpicture}
    \node[anchor=south west,inner sep=0] (image) at (0,0) {\includegraphics[width=\textwidth]{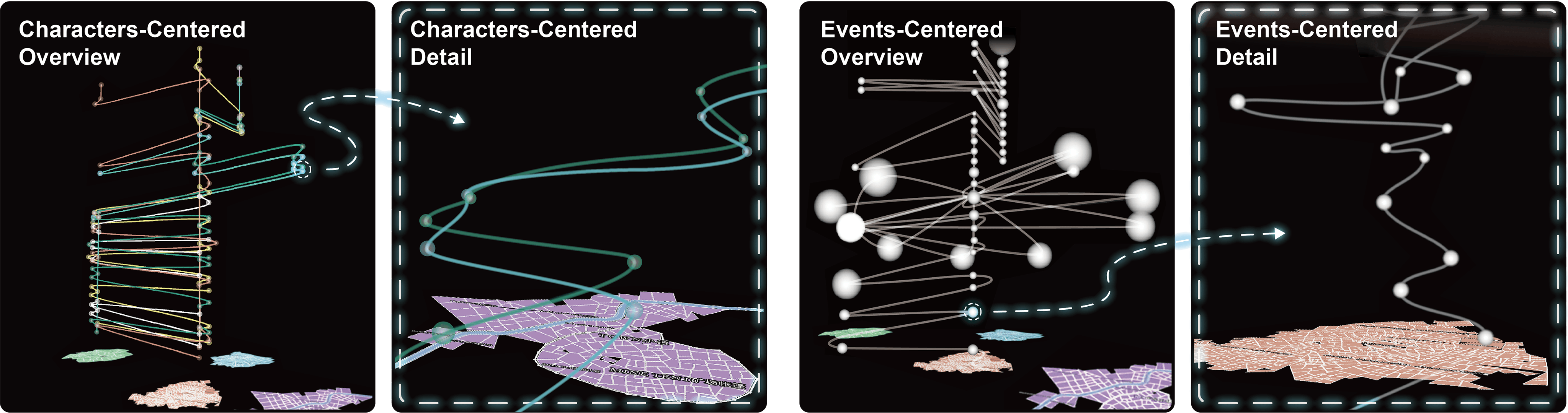}};
    \begin{scope}[x={(image.south east)},y={(image.north west)}]
    \node[font=\large, text=white] at (0.03,0.07) {(a)};
    \node[font=\large, text=white] at (0.275,0.07) {(b)};
    \node[font=\large, text=white] at (0.53,0.07) {(c)};
    \node[font=\large, text=white] at (0.785,0.07) {(d)};
    \end{scope}
  \end{tikzpicture}
  \caption{Storyline visualization of the TV episode story \textit{Loki} in VR environment. Users can switch between narrative perspectives to focus on either a characters-centered or events-centered view: (a) Characters-Centered Perspective in Overview, (b) Characters-Centered Perspective in Detail, (c) Events-Centered Perspective in Overview, and (d) Events-Centered Perspective in Detail.}
	\label{fig:teaser}
}

\vgtcinsertpkg

\begin{document}



\maketitle

\input{body/1_Introduction}

\input{body/2_RelatedWork}

\input{body/3_DesignConsideration}
\input{body/4_3DStoryline}
\input{body/5_Evaluation}
\input{body/6_Discussion}
\input{body/7_Conclusion}

\acknowledgments{
This work was supported by the National Natural Science Foundation of China (Grant No. 62272396).
}

\bibliographystyle{abbrv-doi-narrow}

\bibliography{Reference-output}

\end{document}

%% file: body/1_Introduction.tex
\section{Introduction}
\label{sec:Introduction}
The last decade has seen a concentrated research focus on storyline visualization, an effective visual representation demonstrating the evolution of time-series datasets. 
Initially, it was used to illustrate a movie narrative, with each character denoted by a single line~\cite{munroe2009movie}.
The character interactions are conveyed by the convergence and the divergence of two corresponding lines at an instant.
Typically, one axis encodes time from left to right, and the other encodes a value in these visualizations, aiding viewers in comprehending temporal patterns of entity relationships~\cite{liu2013storyflow}.
In an attempt to balance comprehensive information presentation, aesthetic appeal, and visualization compactness, researchers have developed various optimization methods for visualization algorithms. 
Qiang \etal ~\cite{qiang2017storytelling} conveyed a greater amount of story information using hierarchy and sector mapping. 
Gronemann \etal ~\cite{gronemann2016crossing} computed and leveraged a layout that has the minimum number of line crossings to increase the legibility and aesthetics.

Most storyline-embedded applications and tools developed so far utilize 2D visualizations on traditional surfaces. While this method effectively depicts entire stories, it offers limited space and dimensions to represent complex temporal relationships among numerous entities. Often, the presence of many lines leads to issues such as visual clutter and obscured visualization, making it challenging to follow the lines and understand the events.
To address these issues, some previous work has deviated from traditional design principles and optimization goals, such as avoiding line wiggles, line crossings, and white space. For instance, Tang et al.\cite{tang2018istoryline} conducted several user studies and interviews, discovering that users preferred utilizing white space and line features to present narratives in hand-drawn storylines. This finding motivates us to explore broader spaces for presenting storylines, such as adding an extra dimension to the visualization and presenting storyline visualizations in immersive environments.

In recent years, the rapid development of immersive technologies like virtual/augmented reality (VR/AR) has captured the attention of researchers in the field of visualization.
Unlike the traditional monoscopic 2D monitor screen, immersive environments provide stereoscopic rendering and allow users to interact with data visualizations intuitively using 6 DOF (degrees of freedom) input.
These technologies offer substantial benefits for the comprehension of data visualizations and task accuracy across various types of datasets, including point clouds \cite{10292508, Franzluebbers:2022:VRP,Kraus:2020:TII}, climate data \cite{helbig:2014:CWV}, geographic data \cite{yang:2018:ODF} and historical fragment visualizations \cite{derksen:2023:TDM}. 
The immersive and engaging features of VR/AR enable users to focus more effectively on data features and tasks related to domain-specific problems. Furthermore, because they offer significant advantages for storytelling by fostering a sense of immersion and empathy \cite{hood2020artificial}, immersive technologies are also highly beneficial in education~\cite{doolani2020vis} and entertainment~\cite{dal2017museum}, enhancing the user experience in these areas. Based on this understanding, we predict that these immersive environments would also be highly engaging for depicting narrative stories, providing users with a deeper connection to the story elements and the progression of the narrative. However, it remains unclear how to best leverage the advantages of immersive environments in storytelling to maximize their potential.

These questions prompted us to investigate effective storytelling techniques in VR, starting with the traditional storytelling technique---storyline visualization. When considering 3D storylines in VR, many factors need to be addressed. First, we need to consider the layout of the storyline: How to construct a storyline in 3D space so that key features or main characters and story shots are easily noticeable by viewers. Time and location of events are two key aspects of stories. Traditional 2D storylines represent time as a sequence of events presented from left to right. Tang et al.\cite{tang2020plotthread} proposed a mixed-initiative approach to support easy customization of storyline visualizations. Therefore, the spatial location of events can also be considered in storyline designs. Although 3D space is a promising environment for presenting spatio-temporal data, it remains unexplored how to best arrange the characters (who), time (when), locations (where), and events (what) within it. Second, through the workshop of iStoryline\cite{tang2018istoryline}, we observed that people are usually attracted to different aspects of a story, such as main characters, significant events, and sometimes minor or emotional events. When creating storylines, these ``important'' narrative perspectives often shape the entire visualization. Thus, it would be interesting to explore approaches that utilize the unique features of immersive environments to organize stories for various narrative perspectives. Lastly, one of the key factors that distinguishes immersive storytelling from traditional 2D stories is that users are situated inside the story. This feature allows for many innovative designs. For instance, users can walk around an event or character, and they can interact directly with these narrative elements. Understanding how to best use these unique features to enable users to comprehensively understand narratives is crucial. 

In this work, we conducted a preliminary user study to understand how users interpret narratives in VR. We then proposed design considerations for 3D storylines from the following aspects: narrative perspectives, visual encodings, storyline visualization, layout optimization, and user interaction.
Based on these considerations, we developed 3DStoryline, a novel visualization and interaction tool for immersive storytelling. To evaluate its usability and effectiveness, we conducted a formal user study. The results demonstrated great potential in assisting users to understand complex story structures.
In summary, we make the following three main contributions:
\begin{itemize}[nosep]
\item generating design considerations for 3D storyline visualizations in immersive environments;
\item proposing 3DStoryline, a novel tool that leverages the immersive capabilities of VR to present complex narratives through 3D storyline visualization; and
\item discussing insights and suggestions for future research and development in immersive storytelling techniques, highlighting the potential and challenges of using VR for narrative visualization.
\end{itemize}
\vspace{-.5em}

%% file: body/2_RelatedWork.tex
\section{Related Work}
\label{sec:related}
Our research centers on two main aspects: storyline visualization and immersive storytelling. We first reviewed the related work on these two topics. 
Additionally, our work is highly related to spatio-temporal visualization and incorporates conventions of Space-Time Cube (STC) visualization in the immersive environment. Thus, we review immersive spatial-temporal visualization in the third section.

\subsection{Storyline Visualization}
Since Munroe \cite{munroe2009movie} introduced hand-drawn narratives for the XKCD comics in 2009, storyline visualization has found extensive applications across various fields.
Cui \etal~\cite{Cui:2011:TextFlow} utilized storyline visualization to help readers understand evolving topics in the narrative. 
Lu \etal~\cite{lu2014effective} proposed storyline visualization to demonstrate cooperative relations among the developers in project management. 
Ren \etal~\cite{Ren:2023:ReunderstandingOD} proposed a novel classification scheme for authoring tools based on narrative perspectives.
Storyline visualization is also frequently used to combine geographical information~\cite{he2020integrated,hulstein:2022:Geo-Storylines}, integrating geospatial context into storyline visualizations by employing various strategies to composite time and space.
Progress in narrative visualization, especially in layout algorithms, has promoted research into automating storyline algorithms to balance comprehensive information presentation, narrative aesthetics and computational efficiency.
Ogawa \etal~\cite{Ogawa_Ma_2010} proposed design considerations for line layout from an aesthetic perspective.
Liu \etal~\cite {liu2013storyflow} developed layout optimization algorithms to efficiently visualize hierarchical relationships among entities over time, enhancing the storyline's aesthetic appeal and making it easier to comprehend and follow.
Tanahashi \etal~\cite{tanahashi2015efficient} utilized storyline visualizations for streaming data, enabling users to track and interpret dynamic information effectively.
Research and methodologies have also introduced multi-layer storyline visualization techniques~\cite{padia2019system,padia2018yarn} to showcase various timelines simultaneously, as well as nested layouts for numerous entities~\cite{pena2022hyperstorylines}.

\subsection{Immersive Storytelling}
Immersive storytelling is a narrative technique that creates interactive and engaging experiences where the audience feels as though they are part of the story via 360-degree video~\cite{elmezeny2018immersive,eiris2020safety} or head-mounted display devices~\cite{ceuterick2021immersive}. This technique is extensively utilized across diverse fields, including education~\cite{doolani2020vis}, journalism and communications~\cite{dowling2019immersive}, medical training~\cite{hardie2020nursing}, and museum exhibitions~\cite{dal2017museum}.
The benefits of immersive storytelling have been extensively studied. Immersive environments enhance user immersion and empathy in storytelling ~\cite{hood2020artificial}, facilitating a deeper connection with the characters ~\cite{chopra2021reality, hollick2021work}. Furthermore, interactions within immersive storytelling can significantly boost user engagement~\cite{zhang2019exploring}, which further enhances user motivation ~\cite{mystakidis2014playful}. Natural interactions minimize user distraction, allowing for high-level engagement within the story ~\cite{liang2017exploitation}.
Additionally, interaction can minimize the cognitive load associated with understanding complex narratives. According to narrative theory~\cite{NarrativeTheory}, narrative elements, structure, hierarchy, scale, spatial-temporal frame, and context need to be considered in storytelling. Interactive storytelling can help manage them, preventing them from overwhelming the user simultaneously.
Despite the benefits of interactive storytelling in immersive environments, there is a noticeable lack of systematic research specifically focused on 3D storyline visualization techniques. This includes aspects such as encoding, layout, and interaction design.

\subsection{Immersive Spatio-Temporal Visualization}
Exploring spatio-temporal data is a common task in geovisualization field, which is an interdisciplinary field involving geographic information science, cartography and data visualization~\cite{Kraak:2006:BG}. 
A frequently adopted visualization technique for spatio-temporal data is the Space-Time Cube (STC), which illustrates the data across spatial and temporal dimensions inside a cubic volume, originally introduced in the 1970s~\cite{ilagcrstrand:1970:WPR}. 
Spatio-temporal data visualization is usually leveraged together with a map representation to demonstrate the spatial features~\cite{yang:2019:RVS} and is situated above the map with various visual representations such as trajectories~\cite{tominski:2012:SBV,wagner:2019:evaluating} and bars~\cite{ready:2018:IVB}.
The horizontal surface is often used to represent geographic spaces and the vertical direction is encoded by time.
In recent years, the rapid development of immersive environments has bolstered research on immersive STC~\cite{ens:2020:uplift,wagner:2024:TaxiVis}. 
The inherent 3D feature of the immersive environment greatly amplifies the advantages of STC.
Wagner \etal ~\cite{wagner:2019:evaluating} first introduced STC into the Immersive Analytics domain.
Zhang \etal ~\cite{zhang:2022:timetables} proposed TimeTables to facilitate data exploration on STC with embodied interactions in virtual reality, demonstrating high usability in spatio-temporal data analysis.
3D hexglyph maps~\cite{horst:2022:HexglyphMaps} is a multi-variable data visualization method that merges STC and Hexbin Map techniques within the VR environment, which is proficient in analyzing complex trajectory match data from notable electronic sports games.

Storyline visualization and STC visualization are both techniques for representing and analyzing spatio-temporal data, but they emphasize different aspects and are used in distinct contexts. 
STC focuses on the precise representation of spatio-temporal data and is widely applied in contexts that require exact spatio-temporal analysis. 
In contrast, storyline visualization highlights narrative and character interactions.
Although studies in immersive data analysis have examined STC visualization and interaction, there has been a notable lack of research addressing how to visualize narrative stories and organize spatial layouts from a high-level perspective in the immersive environment.

%% file: body/3_DesignConsideration.tex
\section{Design Consideration}
\label{sec:Design Consideration}

To better understand users’ perspectives in depicting narratives in VR, we conducted a preliminary user study with 12 participants from our local university. Based on the findings, we formulated design considerations for creating immersive storyline visualizations.
\begin{figure}[t]
  \centering
  \begin{tikzpicture}
    \node[anchor=south west,inner sep=0] (image) at (0,0) {\includegraphics[width=\columnwidth]{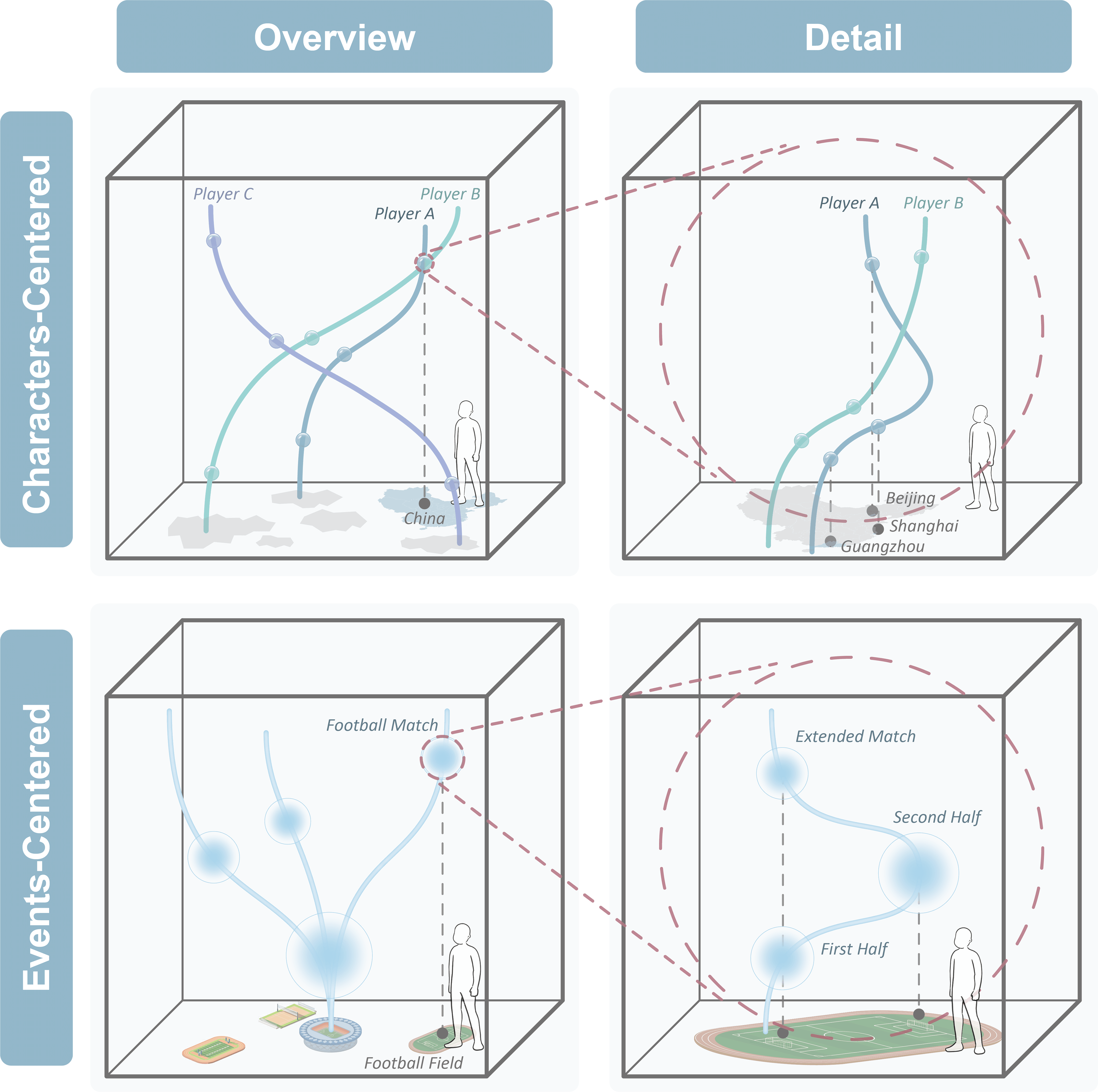}};
    \begin{scope}[x={(image.south east)},y={(image.north west)}]
    \node[font=\large] at (0.50,0.50) {(a)};
    \node[font=\large] at (0.50,0.02) {(c)};
    \node[font=\large] at (0.97,0.50) {(b)};
    \node[font=\large] at (0.97,0.02) {(d)};
    \end{scope}
    \end{tikzpicture}

  \caption{The narrative perspectives. The vertical upward direction encodes time while the horizontal surface represents the geographical map.}
  \label{fig:Schemes}
\end{figure}
\subsection{Preliminary User Study}
\label{subsec:preliminary}
This section introduces the preliminary study and our findings. 

\textbf{Participants.}
We invited 2 domain experts in the field of visualization from the local university, each with \textgreater3 years of professional experience in visual storytelling design and development. 
In addition, we recruited 10 unpaid participants (5 male, 5 female), 20--32 years old, to share their suggestions. Among them, 4 use VR at least once a week, 4 at least once a year, and 2 have never used VR devices. On average, they owned on average 3.7 years of experience in the visualization field.

\textbf{Task and Procedure.}
We began by introducing the basic concept of storyline visualization to the participants. We displayed two 2D storyline visualizations for the movies \textit{Jurassic Park} and \textit{The Moon and Sixpence}, using iStoryline \cite{tang2018istoryline} and StoryFlow \cite{liu2013storyflow} illustrations. We explained how visual elements were used to represent narrative components. Once participants had a basic understanding of storyline visualizations, we demonstrated a story piece selected from the VR game \textit{Half-Life: Alyx} to show alternative ways to depict stories in VR.

We then invited participants to revisit storyline visualizations, this time in VR, and consider the following questions: 

\begin{itemize}[nosep]
\item How should these storylines look in VR (layout)?
\item What visual elements would best represent events and characters (visual elements)?
\item What aspects of the story would they focus on if presented in VR (narrative aspects)?
\item How would they want to interact with the visual/narrative elements (interaction)?
\end{itemize}

Participants were encouraged to share their insights at any time. Afterward, we conducted semi-structured interviews to collect ideas.

\textbf{Findings.}
Participants mentioned diverse narrative perspectives they would focus on if the storyline were presented in VR. For instance, they would like to follow the lines (characters) to understand what the character has been involved in throughout the story. Some participants noted that VR offers a significant advantage in presenting the location and chronological order of events. They could stand in a position (a location in the story) and view what events have happened there.

From their responses, we noticed that characters and events remain the most important narrative perspectives. This finding aligns with the design of traditional 2D storylines and is consistent with Schell's conceptualization of storytelling, which emphasizes the importance of characters and events \cite{schell:2008:artGD}. Participants expressed interest in knowing the characters involved in specific events, the relationships between multiple characters (interaction among characters), and the sequence of events. The layout of the storyline may depend on the narrative perspective.

For conveying this information, line visualizations would still be effective in VR to connect events over time. Points could be used to represent the spatial-temporal position of the character, and spheres could be used to mark key events (similar to closed contours filled with color in 2D storyline~\cite{tanahashi2012design}). The spatial positions on the floor can be well-designed to present locations in the story. Additionally, four participants highlighted the importance of focusing on specific events or characters within the context of the story.

\subsection{Design Considerations}
\label{subsec:designconsideration}
Based on this understanding, we formulate the following design considerations for 3D storyline visualization in VR:

\textbf{DC1. Narrative Perspective.} Adapt the layout and visual elements based on the narrative perspective. Consider how characters, events, and their relationships are best represented from different narrative perspectives.

\textbf{DC2. Visual Encodings.} Use line visualizations to connect events over time effectively. Spheres can be employed to mark key events, and various visual elements should be considered to represent characters, events, and their interactions.

\textbf{DC3. Layout Design.} Arrange the storyline in 3D space to ensure that key features, main characters, and significant events are easily noticeable by viewers. Utilize spatial position to represent different locations in the story.

\textbf{DC4. User Interaction.} Design interactions that allow users to explore the storyline intuitively. Enable users to walk around events and characters, and interact directly with narrative elements to gain a comprehensive understanding of the story.

\begin{figure*}[t]
  \centering
  \begin{tikzpicture}
      \node[anchor=south west,inner sep=0] (image) at (0,0) 
      {\includegraphics[width=\textwidth]{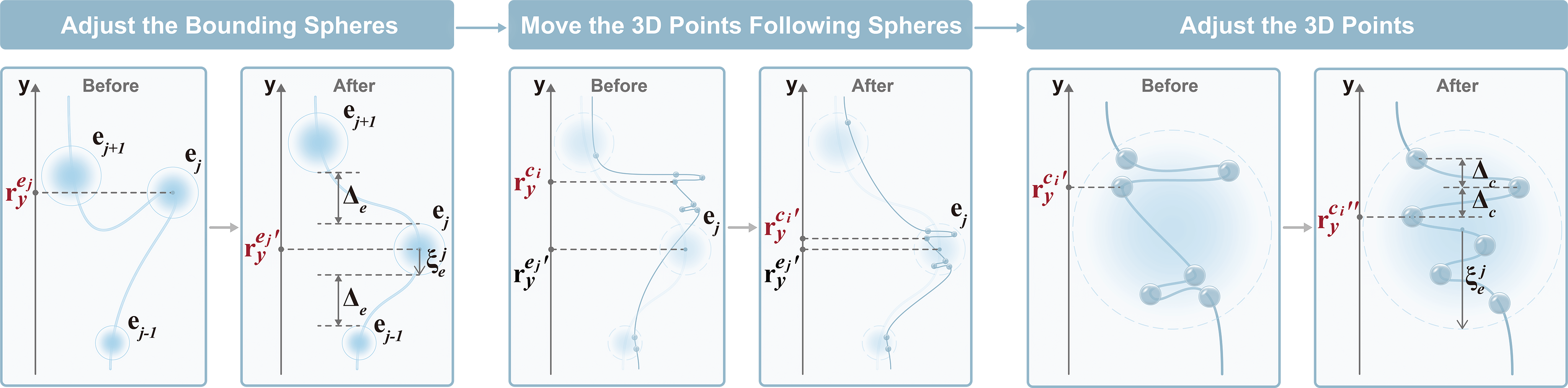}};
      \begin{scope}[x={(image.south east)},y={(image.north west)}]
      \node[font=\large] at (0.015,0.05) {(a)};
      \node[font=\large] at (0.340,0.05) {(b)};
      \node[font=\large] at (0.670,0.05) {(c)};
      \end{scope}
  \end{tikzpicture}
  \caption{The method for optimizing storyline layout along the time axis involves three steps: (a) adjusting the coordinates of bounding spheres to ensure a consistent distance $\Delta_e$ between them, (b) updating the positions of 3D points within each bounding sphere, and (c) fine-tuning the 3D points inside each sphere to maintain a uniform distance $\Delta_c$ between them.}
  \label{fig:Time_Layout_Algrithm}
\end{figure*}

%% file: body/4_3DStoryline.tex
\section{3DStoryline}
\label{sec:3DStoryline}
Based on the design considerations and preliminary study findings, we developed \textbf{3DStoryline}. In this section, we demonstrate design considerations from the following aspects: narrative perspective, visual encodings, storyline visualization, layout optimization, and user interaction. 

\subsection{Narrative Perspective}
\label{subsec:NP}
To provide diverse narrative perspectives \textbf{(DC1)}, We introduce two approaches: Events-Centered and Characters-Centered (as shown in \autoref{fig:Schemes}). 

\begin{description}[nosep,leftmargin=1.5em,labelindent=0em,leftmargin=!,labelindent=!,itemindent=!,font=\normalfont\itshape]
\item[Events-Centered Perspective:] This focuses on the story's development, including the sequence of events (temporal), their locations (spatial), and the correlations between them. Relevant events are connected to indicate their time sequence and relationships, helping readers develop a comprehensive understanding of the story's progression.
\item[Characters-Centered Perspective:] This allows users to follow specific characters, including the events they experience, changes in their locations, and their relationships.
\end{description}

In addition, the preliminary study revealed that users are interested in both the overall story and specific events. Therefore, we support users in obtaining both an ``overview'' of the story and ``detail'' of particular events.

\begin{description}[nosep,leftmargin=1.5em,labelindent=0em,leftmargin=!,labelindent=!,itemindent=!,font=\normalfont\itshape]
\item[Overview:] The entire story, including the time sequence, events, locations, characters and their interactions, is presented. This allows readers to gain a comprehensive understanding of all events and characters.
\item[Detail:] A specific event is presented clearly, showing the location, time and characters involved in that event.
\end{description}

Note that, this ``detail'' searching can be based on location (``what happened here''), time (``what happened at that time''), or character (``what happened to this person''). This allows readers to view both the ``overview'' and ``details'' from both the ``Events-Centered'' and the ``Characters-Centered'' perspectives.

\subsection{Visual Encodings}
\label{subsec:VE}
Following \textbf{DC2}, we use 3D points, bounding spheres, and connecting lines to depict story developments in VR.
\begin{description}[nosep,leftmargin=1.5em,labelindent=0em,leftmargin=!,labelindent=!,itemindent=!,font=\normalfont\itshape]
\item[3D Points:] A 3D point represents the location (at different scales) of a character, visualized only in the \textit{Characters-Centered} design. When the detail view is enabled, more points show detailed locations the character visited. For example, in the overview (\autoref{fig:Schemes}(a)), a 3D point can represent a football player having matches in China. When zooming in, more points appear (\autoref{fig:Schemes}(b)), showing matches in Shanghai, Beijing, and Guangzhou.
\item[Bounding Spheres:] A bounding sphere indicates an event (at different scales), appearing only in the \textit{Events-Centered} design. When the detail view is enabled, more bounding spheres show sub-events. For instance, in the overview (\autoref{fig:Schemes}(c)), a bounding sphere represents a football match. When zooming in (\autoref{fig:Schemes}(d)), more details such as the goals in the first half, second half, and extended match will appear as distinct bounding spheres. The size of the sphere (radius) indicates the event's importance to the entire story.
\item[Lines:] Lines connect points or bounding spheres, indicating the traveling path of characters or the sequence/correlation of events.
\item[Color:] Color is used to differentiate characters in \textit{Characters-Centered} design (\autoref{fig:Schemes}(a) and (b)).
\end{description}

\subsection{Storyline Visualization}
We leveraged the immersive Space-Time   convention~\cite{ens:2020:uplift,wagner:2024:TaxiVis,zhang:2022:timetables}, in which the vertical upward direction is encoded by time and the horizontal plane is leveraged to represent 2D geographic information, to visualize the spatio-temporal narrative data by the Characters- and Events-Centered views proposed in \autoref{subsec:NP}.

\textbf{Characters-Centered View.}
The concept of 3D points mentioned in \autoref{subsec:VE} refers to the character $\mathbf{c}$'s spatial-temporal coordinate extracted from data entries in the original dataset. 
For a character $\mathbf{c}$, we obtained a list of spatial-temporal 3D points from the original dataset as $\mathbf{r}^{c}=\{{\mathbf{r}^{c_0},\mathbf{r}^{c_1}...\mathbf{r}^{c_n}}\}$, in which $(\mathbf{r}_x^{c_i},\mathbf{r}_z^{c_i})$ represents the 2D geo-location and $\mathbf{r}_y^{c_i}$ is the time of the $i^{th}$ 3D point $\mathbf{r}^{c_i}$.
Each 3D point $\mathbf{r}^{c_i}$ has an impact factor $\mathbf{\xi}_c^i$, which dictates whether $\mathbf{r}^{c_i}$ should be visualized in the overview or detail. 
when $\mathbf{\xi}_c^i$ is larger than a pre-defined global threshold $\mathbf{\xi}_{c}^{thre}$, $\mathbf{r}^{c_i}$ is visualized in overview by tiny transparent balls with unique size (\autoref{fig:Schemes}(a)). 
Vice versa for detail, if $\mathbf{\xi}_c^i \leq \mathbf{\xi}_{c}^{thre}$, $\mathbf{r}^{c_i}$ is visualized in detail (\autoref{fig:Schemes}(b)).
Next, in both overview and detail, we link the 3D points by time sequence for each character with lines in different colors by cubic spline interpolation.
To be noted, the 3D points visualizations and the character lines are only visualized in the ``Characters-Centered'' Perspective. 
In the ``Events-Centered'' Perspective, it is hidden.

\textbf{Events-Centered View.}
We collect all the events $\mathbf{e}=\{\mathbf{e}_{0},\mathbf{e}_{1}...\mathbf{e}_{m}\}$ from the dataset. For the $j^{th}$ event $\mathbf{e}_{j}$, it has attributes including (1) the start time $\mathbf{t}_{start}^{j}$, (2) the end time $\mathbf{t}_{end}^{j}$, (3) the spatio-temporal coordinate $\mathbf{r}^{e_j}$, where $(\mathbf{r}_{x}^{e_j},\mathbf{r}_{z}^{e_j})$ is the 2D geo-location where it happens and $\mathbf{r}_{y}^{e_j}=\frac{\mathbf{t}_{start}^{j}+\mathbf{t}_{end}^{j}}{2}$, and (4) event impact factor $\mathbf{\xi}_{e}^j$, which is quantified by how long the event lasts in our design.
when $\mathbf{\xi}_e^j$ is larger than a pre-defined global threshold $\mathbf{\xi}_{e}^{thre}$, $\mathbf{r}^{e_j}$ is visualized in overview by the opaque bounding sphere with the radius equal to $\mathbf{\xi}_e^j$ (\autoref{fig:Schemes}(c)). 
Vice versa for detail, if $\mathbf{\xi}_e^j \leq \mathbf{\xi}_{e}^{thre}$, $\mathbf{r}^{e_j}$ is visualized in detail (\autoref{fig:Schemes}(d)).
Then, we connect the center of the bounding spheres based on the sequence of the events with lines by the cubic spline interpolation.
Be noted that the event visualization and event lines only exist in the ``Events-Centered '' Perspective. 

\subsection{Layout Optimization}
To meet \textbf{DC3}, we develop the layout optimization methods for storyline visualization along the time axis, as well as for 2D geo-map.

\textbf{Storyline Layout along Time Axis.}
The narrative often spans extensive time scales and is distributed nonlinearly along time axis, which can lead to widespread distribution or overlapping of visual elements. To prevent space wastage and visual clutter, we propose a nonlinear mapping method that expands dense regions and compresses sparse ones. This is achieved by rearranging 3D points and bounding spheres along the time axis in three steps, ensuring a predefined distance between all the bounding spheres.
This layout optimization is applied separately to both the overview and detail views.

First, we recalculate the coordinates along the time axis of each bounding sphere. 
For $j^{th}$ bounding sphere, we calculate the new time coordinate ${\mathbf{r}_{y}^{e_j}}^{\prime}$, ensuring that the distance along the time axis between the boundary of event sphere $\mathbf{e}_{j}$ and the neighbor event sphere $\mathbf{e}_{j+1}$ is $\Delta_e$ (\autoref{fig:Time_Layout_Algrithm}(a)), denoted by:
\begin{equation}
    {\mathbf{r}_y^{e_{j+1}}}^{\prime}-{\mathbf{r}_y^{e_{j}}}^{\prime}=\Delta_e+\xi_{e}^{j+1}+\xi_{e}^j,      j \in [0,m-1]
\label{equ:dist_e_def}
\end{equation}
We pre-define the time coordinate of $0^{th}$ sphere ${\mathbf{r}_y^{e_{0}}}^{\prime}$.
We then derive ${\mathbf{r}_y^{e_{j}}}^{\prime}$ from \autoref{equ:dist_e_def} as:
\begin{equation}
    {\mathbf{r}_y^{e_{j}}}^{\prime}={\mathbf{r}_y^{e_{0}}}^{\prime}+\sum_{\tau=0}^{j-1} (\Delta_e+\mathbf{\xi}_{e}^\tau+\mathbf{\xi}_{e}^{\tau+1}),    j \in [1,m]
\end{equation}

Second, we move all the 3D points inside the bounding spheres to the new positions, following the movement of the bounding spheres (\autoref{fig:Time_Layout_Algrithm}(b)).
For a given 3D point $\mathbf{r}^{c_i}$, if
\begin{equation}
    \exists j \in m, \left|  \mathbf{r}^{c_i}-\mathbf{r}^{e_j} \right| \leq {\xi_e^j}
\end{equation}
, we say the character $\mathbf{c}$ is involved in the event $\mathbf{e}^{j}$ at 3D point $\mathbf{r}^{c_i}$ at moment $\mathbf{r}_y^{c_i}$ with geo-location $(\mathbf{r}_x^{c_i},\mathbf{r}_z^{c_i})$.
We denote this relationship by $\mathbf{r}^{c_i} \in \mathbf{e}_{j}$.
Then we move the 3D point $\mathbf{r}^{c_i}$ to the new coordinate ${\mathbf{r}^{c_i}}^{\prime}$ based on the movement of $\mathbf{e}_{j}$'s bounding sphere:
\begin{equation} 
    {\mathbf{r}_y^{c_i}}^{\prime}=\mathbf{r}_y^{c_i}+({\mathbf{r}_y^{e_{j}}}^{\prime}-\mathbf{r}_y^{e_{j}})
\end{equation}

\begin{figure}[t]
  \centering
  \begin{tikzpicture}
      \node[anchor=south west,inner sep=0] (image) at (0,0) 
      {\includegraphics[width=\columnwidth]{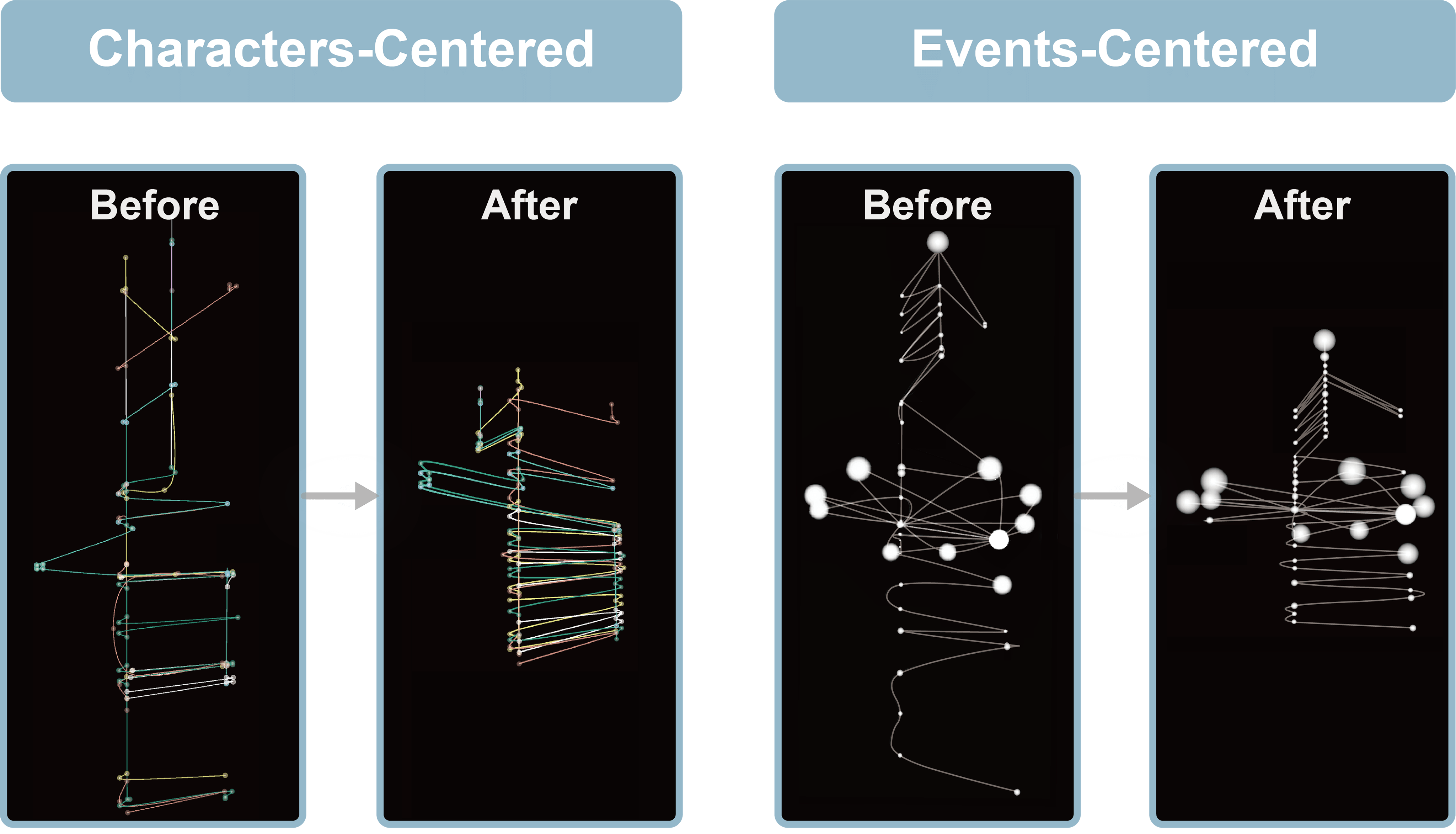}};
      \begin{scope}[x={(image.south east)},y={(image.north west)}]
      \node[font=\large, text=white] at (0.035,0.06) {(a)};
      \node[font=\large, text=white] at (0.567,0.06) {(b)};
      \end{scope}
  \end{tikzpicture}
  \caption{Result of storyline layout optimization along time axis: (a) Characters-Centered Perspective and (b) Events-Centered Perspective.}
  \label{fig:CompareAlgorithm}
\end{figure}

Third, we rearrange the 3D points inside the event sphere to a sequential layout~\cite{7581076} (\autoref{fig:Time_Layout_Algrithm}(c)) with a uniform distance $\Delta_c$ between ${\mathbf{r}_y^{c_i}}^{\prime}$ and ${\mathbf{r}_y^{c_{i+1}}}^{\prime}$ along the time axis, where $\Delta_c$ is defined by:
\begin{equation}
   \Delta_c=\frac{2\xi_e^j}{N}
\label{equ:dist_c_def}
\end{equation}
For a given bounding sphere of event $\mathbf{e}_{j}$,
if a 3D point ${\mathbf{r}^{c_i}}^{\prime} \in \mathbf{e}_{j}$, we update it's time coordinate to ${\mathbf{r}^{c_i}}^{\prime\prime}$ by:
\begin{equation}
    {\mathbf{r}_y^{c_i}}^{\prime\prime}={\mathbf{r}_y^{e_{j}}}^{\prime}+(\mathbf{Index}-\frac{N}{2})*\Delta_c
\end{equation}
where ``N'' is the number of 3D points enclosed by sphere of event $\mathbf{e}_{j}$ and ``Index'' is 3D point ${\mathbf{r}^{c_i}}^{\prime}$'s index among all the enclosed 3D points.  
By following three steps, we achieved a uniform layout along the time axis (\autoref{fig:CompareAlgorithm}).

\begin{figure}[t]
  \centering
  \begin{tikzpicture}
      \node[anchor=south west,inner sep=0] (image) at (0,0) 
      {\includegraphics[width=\columnwidth]{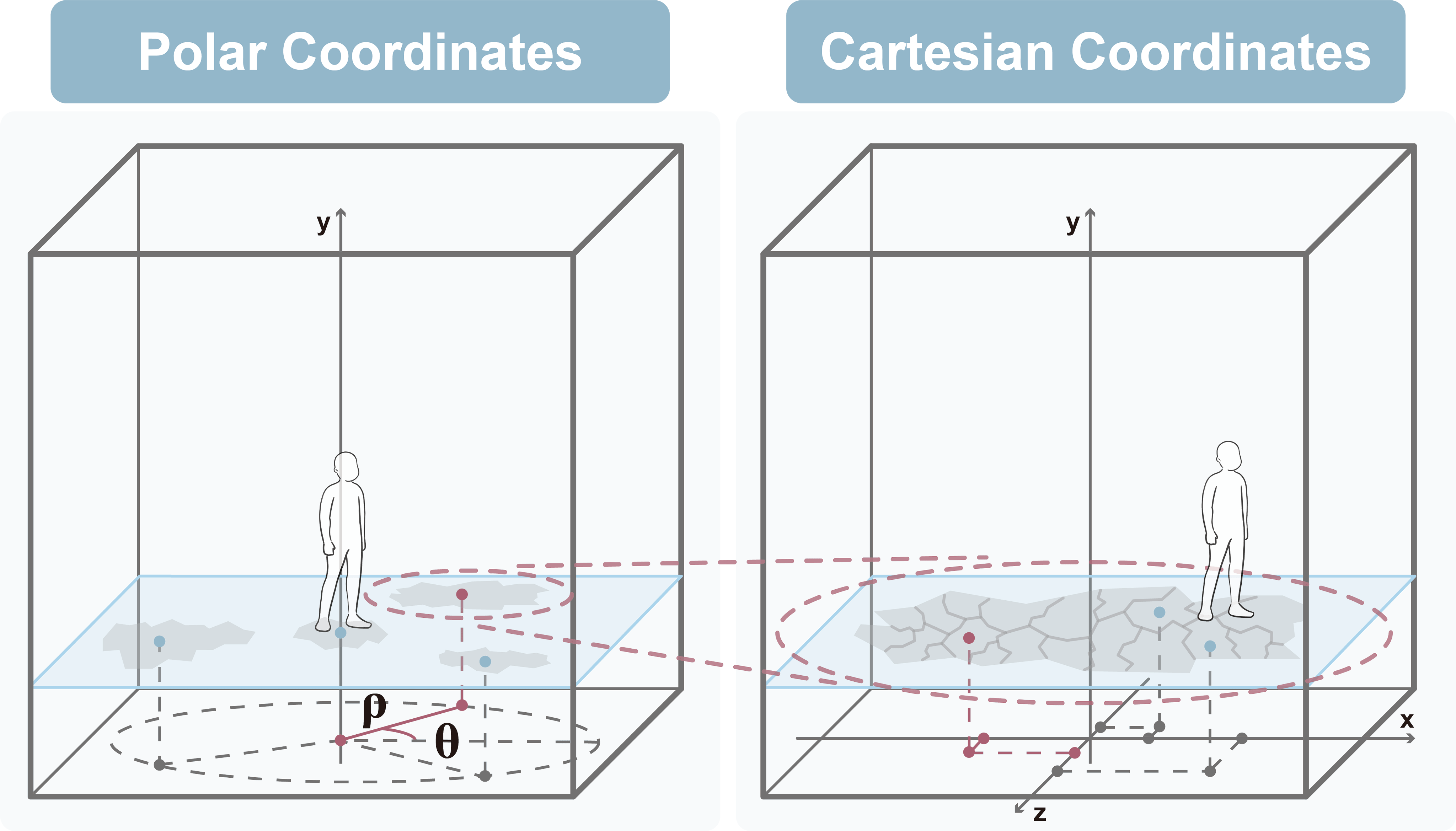}};
      \begin{scope}[x={(image.south east)},y={(image.north west)}]
      \node[font=\large] at (0.46,0.06) {(a)};
      \node[font=\large] at (0.96,0.06) {(b)};
      \end{scope}
  \end{tikzpicture}
  \caption{Geo-map layout in (a) Overview and (b) Detail.}
  \label{fig:AutoLayout}
\end{figure}

\textbf{Geo-map Layout.}
Typically, a narrative story encompasses a variety of scenarios. Characters interact with each other in different scenarios.
Events in different scenarios are correlated. In our design, each scenario is visualized via 2D geographical maps in the immersive environment. 
In detail view, the geo-map of a specific scenario is visualized in Cartesian Coordinates (\autoref{fig:AutoLayout}(b)).
In the overview, multiple geo-maps of the scenario are integrated, allowing the user to gain a comprehensive view of the story. 
We utilize a polar coordinate system centered on the user's position to determine the location of each map ($\rho, \theta$)(\autoref{fig:AutoLayout}(a)). To maintain awareness of key characters and events, maps of higher importance are visualized with a smaller $\rho$. The importance of a map is calculated by integrating the impact factor of each event and the 3D point associated with it. The parameter $\theta$ for each map is calculated using force-directed methods to achieve an optimal layout.
Please note that the layout of the maps in the overview is generated only once, based on the user's initial location when the program starts. It does not update as the user navigates through the immersive environment.

\subsection{User Interaction}
To meet \textbf{DC4}, we develop interaction techniques to (1) switch between Characters-Centered and Events-Centered Perspective, (2) switch between overview and detail mode, (3) navigate, and (4) explore the storyline visualization.

\textbf{Switching. }Specifically, users can use two pre-defined buttons on the VR controller to switch between the Characters-Centered and Events-Centered Perspectives, as well as between the overview and detailed views. When changing perspectives, the visualization of the current perspective gradually fades out while the visualization of the target perspective gradually emerges. When switching from overview to detail, the user will enter the map nearest to their current location.

\textbf{Navigation. }On the one hand, users can either teleport or walk around on the geo-map (ground) within the immersive environment. On the other hand, they can control the storyline, moving it up and down along the time axis, using the sticker/pad on the VR controller.

\textbf{Storyline Exploration. }When the user touches the 3D point, event sphere or the character/event line, basic information (e.g., name, geographic information, time) will be displayed. In addition, we support users in viewing the Character-Centered Perspective view locally enclosing in the bounding sphere of an event by selecting the corresponding bounding sphere. This helps users to explore the correlation between characters and a specific event.

\subsection{Implementation}
we developed 3DStoryline with Unity3D (as shown in \autoref{fig:teaser}). 
We leveraged an all-in-one VR head-mounted display Meta Quest 2 ($1832\times1920$ resolution per eye, $100$° FOV, $90$Hz refresh rate). It was streamed on a PC (Intel Core\texttrademark{} i7 $2.21$GHz, $32$GB RAM, GeForce RTX $3070$, $24$GB video memory).
3DStoryline allows users to load datasets in XLSX or CSV format containing names of characters and events, time-space information, sequence/correlation of events, impact factors of characters and events, and additional attributes displayed in the interaction.

%% file: body/5_Evaluation.tex
\section{Evaluation}
\label{sec:Evaluation}

To evaluate the effectiveness of our design and user experience, we conducted a study using 3DStoryline in a VR environment. Specifically, we aimed to assess our design on narrative perspective, visual encodings, layout design, and user interaction.

\subsection{Dataset and Tasks}
We created a narrative dataset derived from the TV episode story \textit{Loki}, featuring multiple characters traveling across time and various universes. The dataset contains 411 character and 95 event records, with attributes such as start and end times, the importance of characters or events. The dataset is open-source and available at \href{https://github.com/NarrativeDataset/NarrativeDataset_LokiEpisode1}{\texttt{github.com\discretionary{/}{}{/}NarrativeDataset\discretionary{/}{}{/}NarrativeDataset\_LokiEpisode1}}.
In the study, we defined four tasks, each with unique scenarios from the story:

\begin{itemize}[nosep,leftmargin=10pt]
    \item \textbf{Task1.} Finding specific story scenes to check if our tool supports navigating to different detailed scenarios of the story.
    \item \textbf{Task2.} Pointing out specific events or characters to demonstrate the validity of visual encoding for events and characters.
    \item \textbf{Task3.} Identifying a set of events or characters in a specific relationship to evaluate the tool's ability to show relationships between characters or events in detailed narratives.
    \item \textbf{Task4.} Determining the key event in the story that caused a change in a specific character's relationship to show how our tool helps in understanding the story overview.
\end{itemize}

\begin{figure}[t]
  \centering
  \includegraphics[width=\columnwidth]{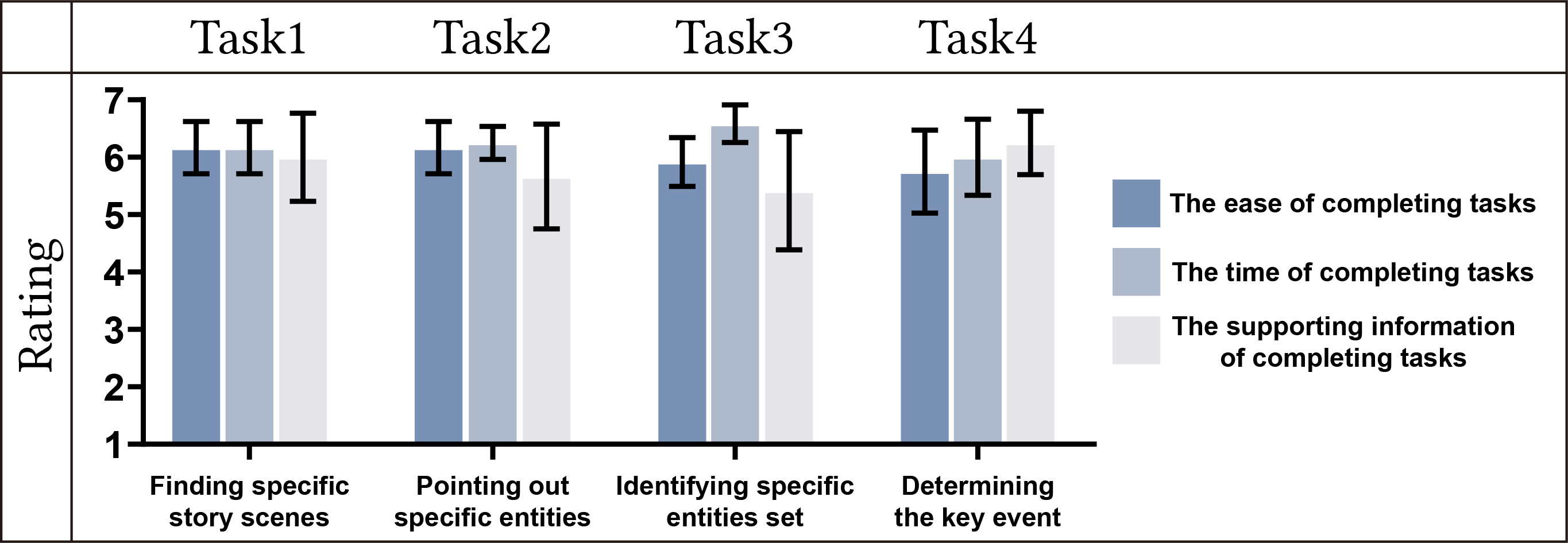}
  \caption{The geometric mean scores of user satisfaction with ease, time, and supporting information from ASQ. Error bars show 95\% confidence intervals.}
  \label{fig:result1}
  \vspace{-1em}
\end{figure}

\subsection{Participants}

We recruited 12 unpaid participants (7 males and 5 females) from a local university, all right-handed, aged 18-28 years old ($Mean$=$24$, $SD$=$3.2$).
All participants majored in computer-related fields, with 10 having a Bachelor’s degree or higher.
All participants had normal or corrected-to-normal vision and could distinguish colors required by our application. Eight participants have previous experience with 2D storytelling systems. None were familiar with the dataset or the TV episode \textit{Loki}.

\subsection{Procedure}
Before the study began, we surveyed participants about their familiarity with the TV episode \textit{Loki}. We then introduced the design of 3DStoryline, explaining the narrative perspective, visual encodings, layout, and interaction techniques. Participants were provided with an example story scenario to explore using our tool until they were comfortable with its design. Afterward, we gave them a brief summary of the \textit{Loki} story and showed a short video about the background and main characters. Then, participants were asked to perform the four tasks in a fixed order.

After each task, participants completed an After-Scenario (ASQ) Questionnaire with 7-point Likert scales. At the end of the study, they completed a System Usability Scale (SUS) questionnaire with 5-point Likert scales, a Presence (PQ) Questionnaire with 7-point Likert scales, and a User Experience Questionnaire (UEQ) with 7-point Likert scales. A semi-structured interview followed to gather additional feedback on tool effectiveness and user experience. Each session lasted approximately one hour.

\subsection{Methodology}
The ASQ measured satisfaction with ease, time, and supporting information for each task. The SUS assessed tool usability, with the final score derived from weighted 7-point scale ratings. Questions from four PQ subscales suitable for immersive environments measured immersion. The UEQ assessed overall user experience, converted from a -3 to 3 scale to a 1 to 7 scale. Quantitative data from the questionnaires were calculated and reported as means and 95\% confidence intervals. Qualitative feedback and behavior observations were organized and classified for presentation.

\subsection{Quantitative Results}
\textbf{ASQ results.}
The ASQ data is shown in \autoref{fig:result1}. 
In \textit{Task1}, satisfaction with the ease ($M$=$6.167$, $CI$=[$5.757$, $6.667$]), time ($M$= $6.167$, $CI$=[$5.757$, $6.667$]), and supporting information ($M$=$6.0$, $CI$=[$5.275$, $6.81$]) had high ratings. 
In \textit{Task2}, satisfaction with the ease ($M$=$6.167$, $CI$=[$5.757$, $6.667$]) and time ($M$=$6.25$, $CI$=[$6.005$, $6.58$]) received high ratings, while supporting information ($M$=$5.667$, $CI$=[$4.797$, $6.622$]) was slightly lower. 
In \textit{Task3}, satisfaction with the ease ($M$= $5.917$, $CI$=[$5.537$, $6.387$]), time ($M$=$6.583$, $CI$=[$6.303$, $6.958$]), and supporting information ($M$=$5.417$, $CI$=[$4.432$, $6.492$]) showed a similar trend. 
In \textit{Task4}, high ratings were achieved for ease ($M$=$5.75$, $CI$=[$5.07$, $6.515$]), time ($M$=$6.0$, $CI$=[$5.38$, $6.705$]), and supporting information ($M$=$6.333$, $CI$=[$5.823$, $6.928$]).
Overall, participants were highly satisfied with the ease and time required to use the prototype to explore narratives. However, satisfaction with supporting information in \textit{Task2} and \textit{Task3} was relatively low, with slightly higher rating uncertainty.

\begin{figure}[t]
  \centering
  \begin{tikzpicture}
    \node[anchor=south west,inner sep=0] (image) at (0,0) 
      {\includegraphics[width=\columnwidth]{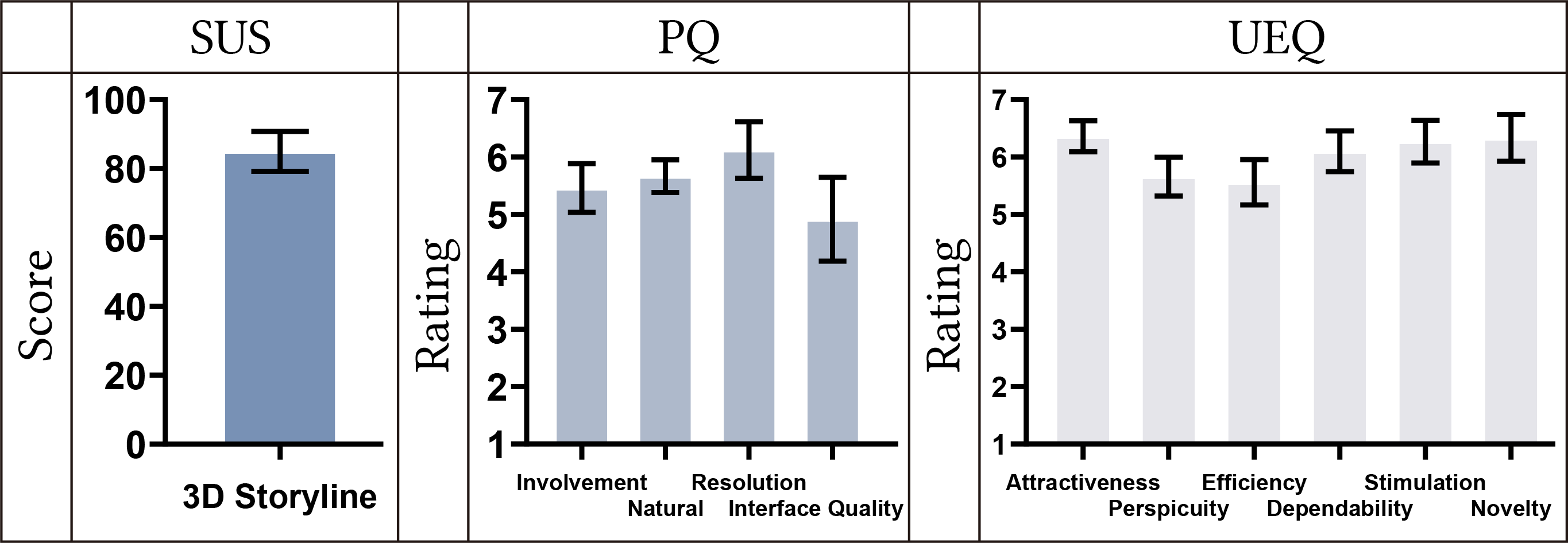}};
      \begin{scope}[x={(image.south east)},y={(image.north west)}]
      \node[font=\large] at (0.023,0.07) {(a)};
      \node[font=\large] at (0.277,0.07) {(b)};
      \node[font=\large] at (0.602,0.07) {(c)};
      \end{scope}
  \end{tikzpicture}
  \caption{The geometric mean score of (a) SUS, (b) PQ, and (c) UEQ. Error bars show 95\% confidence intervals.}
  \label{fig:result2}
\end{figure}

\textbf{SUS results.}
The left chart in \autoref{fig:result2}(a) shows the participants’ SUS final scores for our tool. According to Bangor's adjective rating scale~\cite{bangor2009determining}, the mean of SUS scores ($M$=$85.0$) was near the excellent range, and the confidence interval ($CI$=[$79.916$, $91.499$]) is above the good score, indicating that the prototype was acceptable. And the sample standard deviation ($SD$=$9.108$) is reasonable.

\textbf{PQ results.}
The middle chart in \autoref{fig:result2}(b) illustrates the PQ ratings, divided into involvement, naturalness, resolution, and interface quality. The involvement ($M$=$5.462$, $CI$=[$5.082$, $5.932$]), naturalness ($M$=$5.667$, $CI$=[$5.427$, $5.997$]), and resolution ($M$=$6.125$, $CI$=[$5.675$, $6.660$]) received high ratings, while interface quality ($M$=$4.917$, $CI$=[$4.232$, $5.692$]) was relatively low.
Overall, participants had a good sense of presence when completing tasks using our prototype, but the interface quality sightly negatively impacted their experience.

\textbf{UEQ results.}
The middle chart in \autoref{fig:result2}(c) illustrates the UEQ ratings, which contains six aspects: attractiveness, perspicuity, efficiency, dependability, stimulation, and novelty. Attractiveness ($M$=$6.363$, $CI$=[$6.138$, $6.678$]), dependability ($M$=$6.104$, $CI$=[$5.794$, $6.504$]), stimulation ($M$=$6.271$, $CI$=[$5.941$, $6.686$]), and novelty ($M$=$6.333$, $CI$=[$5.973$, $6.788$]) had high scores. However, perspicuity ($M$=$5.660$, $CI$=[$5.365$, $6.040$]) and efficiency ($M$=$5.563$, $CI$=[$5.213$, $6.002$]) were relatively low.
Overall, our prototype demonstrated good dependability and novelty, and was found to be attractive and stimulating to participants. However, there is potential for improvement in perspicuity and efficiency.

\subsection{Qualitative Feedback}
\textbf{Narrative Perspective.}
All participants easily understood the rationale behind the two narrative perspective designs, \textit{Events-Centered} and \textit{Characters-Centered}. To answer the questions, they could switch between these perspectives easily to view the story from different angles. Additionally, they used both the \textit{Overview} and \textit{Detail} views to explore the broader context or obtain more specific information.

Regarding the time spent on the two perspective views, we noticed that most participants (10 out of 12) frequently toggled between them, searching for answers from both character and event perspectives. Two participants primarily used the \textit{Characters-Centered} view, switching to the \textit{Events-Centered} view only for event-related tasks. They mentioned enjoying the focus on characters as it helped them deeply understand character changes and imagine how these characters influenced the story's development. They noted that traditional storytelling techniques do not allow such a focus on specific characters.

Additionally, the flexibility of switching between \textit{Overview} and \textit{Detail} was highly appreciated. Participants (5 \texttimes) mentioned that both the context and details are equally important for understanding a story. The simple switches between \textit{Overview} and \textit{Detail} and between \textit{Events-Centered} and \textit{Characters-Centered} perspectives significantly supported their ability to quickly comprehend the story background, details, and character relationships.

\textbf{Visual Encodings.}
Visual encodings were easy to understand in the study. Participants noted that they could easily track location changes in the Characters-Centered storylines. The size of the bounding sphere helped them identify important events. The multi-scale design supported their understanding of the overall story trend and the characters' travel paths while allowing them to focus on the details of significant events and actors. Although color was used only to indicate different actors in the study, participants suggested it could be an effective visual cue for indicating additional story information, such as enemies or opponents, character status, or relationships, depending on the story's content.

\textbf{Layout Design.}
The spatial-temporal layout provided a 3D view of location and event timelines. Participants were impressed by the ability to ``stand'' inside the story. This situated visualization helped them understand location, time, events, and actors in various ways. For example, they could focus on a specific city on the map to see which characters visited and when, or what events happened there and in what order. Participants could also follow the \textit{Events-Centered} storyline to grasp the story's progression. One participant stated, ``The map and the storylines support me in better understanding the relationship between characters along the temporal dimension and the correlation between the events and the geographic information.'' Another participant commented, ``Standing in the position of the storyline of the actor, I can feel more about what he was experiencing.'' These findings indicate that 3DStoryline and the spatial-temporal layout effectively support users in understanding ``what happened here,'' ``what happened at that time,'' and ``what happened to this person.''

\textbf{User Interaction.}
In the study, participants quickly understood and remembered the interaction methods. All participants agreed that these methods were easy and effective for detecting story pieces and navigating through the story. However, our main focus was on the storyline visualization in the 3D space, so the interaction methods provided were limited. Participants suggested many innovative interaction designs. For instance, they recommended including a transition when switching between narrative perspectives to better relate the context of the event to the characters involved.

%% file: body/6_Discussion.tex
\section{Discussion}
\label{sec:Discussion}
In this section, we discuss our insights into designing 3D storytelling techniques and presenting storyline visualizations in immersive environments. We also address the limitations of our work and provide suggestions for future developments in immersive storytelling techniques.

\subsection{3D Storytelling Techniques}
\label{subsec:3DStorytelling}

Storytelling techniques have been extensively discussed in previous research~\cite{Ren:2023:ReunderstandingOD}. Stories can be narratives~\cite{mitchell2003review}, movies~\cite{glebas2012directing}, or data-driven stories~\cite{stolper2018data} that convey messages such as findings, insights, or trends through data. These data are collected and visualized to form story pieces, which are then connected in meaningful order to support the original story message. Thus, a story does not need to be limited to 2D. If the data is 3D spatial data or multiple dimensions are required to support the story message, a 3D story may be more effective.

This work focuses on one particular storytelling technique---storyline visualization, which uses lines to represent characters, events, and their relationships. Although we have not compared 3D storylines with 2D ones regarding user preference and story comprehension, we have already seen their potential. Compared with a 2D layout, 3D storylines involve an additional dimension to present information. This raises the question of what this extra dimension adds to storytelling. In this work, we explored the combination of geographical visualization and storyline visualization, situating the storylines based on location of events and characters. Building on this, we organized the stories from two perspectives: \textit{Events-Centered} and \textit{Characters-Centered}. Together, the design effectively presents the key factors of stories: who, where, when and what. This approach offers users a more comprehensive understanding of the story. However, there are many other alternatives for 3D storyline visualizations. For instance, we could combine network visualization with storylines to not only show relationships between individuals over time but also allow us to observe how these relationships change through time.

However, compared to 2D layouts, many design decisions need to be made for 3D storylines. In 2D storylines, data such as changes in individual locations over time are mapped to line curves, trends, and positions, which can be accurately observed on a 2D interface. In contrast, 3D layouts allow users to perceive data from different angles, which can lead to misunderstandings or even ignoring the line curves and the events they represent. Therefore, explicit visual cues indicating the positions of location changes or events become essential. For example, in our work, we use 3D points and bounding spheres to highlight these changes, ensuring that participants can easily notice this information when looking at the 3D stoylines. Therefore, we found that participants were satisfied with the ease and time required to use our tool to explore narratives. 

Moreover, we need to consider the layout and positions of multiple characters and events (lines with 3D points or bounding spheres). Our goal is to ensure that the arrangements of these visual elements are meaningful without being too dense or too sparse, maintaining both aethetics and clarity. However, this distribution is influenced by the time and location distribution of the story pieces or events, as well as the narrative perspectives. In our work, the positions of the storylines are determined by the actual locations of events and characters over time. We carefully designed the spatial distribution of points and bounding spheres along the time axis. 

\subsection{Storytelling in Immersive Environments}
\label{subsec:ImmersiveStory}
Virtual reality and augmented reality provide immersive experiences that help users understand visualization data~\cite{dwyer2018immersive}. In such environments, users feel as if they are inside the data feature. Moreover, if multiple users are involved, they feel as though they are experiencing the data together. In this work, we explored users' experiences of viewing not just simple data  or scenes, but meaningful story visualizations. Here, we share our insights on visual storytelling in immersive environments. 

When designing storytelling techniques for immersive environments, we need to consider how best to utilize the advantage of its features to enable users to comprehensively understand narratives. For instance, VR environment can significantly enhance users' engagement in the story~\cite{mills2021engagement}. In this work, we make use of 3D space to present the spatial-temporal feature of narratives. We allowed users to observe narratives from two narrative perspectives: \textit{Events-Centered} and \textit{Characters-Centered}. Users are fully immersed in the storylines and engaged in exploring events and characters with different interaction methods, including \textit{Overview} and \textit{Detail}, and navigation. The study results indicated that participants had a strong sense of presence. They reported enjoying the feeling of standing inside the story and experiencing events from the actor's perspective. Therefore, we suggest future research can consider using large-scale visualization and situated visualization for immersive storytelling. Immersing users in the narrative and visual elements in VR can effectively engage them in the story.

Through our exploration, we also noticed that interaction techniques are crucial for exploring stories in VR, especially when offering various narrative perspectives. Participants frequently switched between \textit{Events-Centered} and \textit{Characters-Centered} Perspectives (mode changing). They usually observed the story's overview to understand the overall trend, then walked (navigation) to regions of interest (ROI), such as an event or actor, and zoomed in to check more details (overview and detail). Although not yet implemented in the current prototype, we imagine that users would be interested in using their bodies to interact with narrative elements. For instance, crouching to check more details related to a location, gazing to show more details of the focused element, or grabbing lines together to rearrange the storylines. In addition, multimodal interactions, including gestures, voice commands, and even haptic feedback, could provide a richer and more engaging experience. These interactions allow users to engage with the story in a more natural and intuitive manner, enhancing their immersion and understanding.

\subsection{Limitations and Future Work.}
In this work, we explored the possibility of transcending traditional storyline visualization approaches by utilizing 3D structures to depict narratives. Furthermore, we investigated the potential of VR environment in narrative storytelling. In this section, we discuss the limitations of our work and lessons learned. We hope that our efforts will inspire more in-depth research into immersive storytelling.

First, we provide two narrative perspectives (events and characters) and two view modes (overview and detail) in the prototype. To comprehensively understand relevant storylines and viewpoints, users can switch between these perspectives and view modes. However, users may get confused about these two perspectives during narrative exploration. For instance, a user focusing on detailed information about an important event may want to check the actors involved. They can switch from the \textit{Events-Centered} view to the \textit{Characters-Centered} view. In the \textit{Events-Centered} view, bounding sphere explicitly indicated individual events. After switching to the \textit{Characters-Centered} view, the bounding sphere disappears, and involved actors' travelling paths during the event appear. Since the visual designs in both perspectives are similar (points, bounding spheres, and connecting lines), users may get confused about which perspective is enabled. Therefore, future work should extensively consider all perspectives, the visual element designs, and the transitions between them.

Second, the interaction methods provided in the prototype are limited. As discussed earlier, future work could support users in exploring stories using multimodal interactions, such as gestures, gaze, and voice commands. Additionally, the VR environment holds promise for storytelling to a group of audiences, enabling shared experiences and collaborative exploration of narratives. However, many design decisions need to be considered in collaborative storytelling. For instance, what visual content should be shared with all audiences and how they can share their insights. This includes determining the best ways to display shared information, facilitating communication between users, and ensuring that the interactive elements are accessible and engaging for everyone involved. 

Third, we have not yet compared 3D storyline visualizations with traditional 2D designs. The advantages of 2D designs have been extensively explored in previous research, showing that users can comprehend the development of entire stories at a glance. Line features have been effectively mapped to the relationship or characteristics of the actors. On the other hand, we have also noticed the advantages of 3D storylines, which can integrate more information, more perspectives and embed more features in the lines. Future work can further compare these two representation approaches to evaluate their effectiveness, usability, and impact on user comprehension and engagement. 

Last but not least, our work focuses on a specific storytelling approach---storyline visualization. However, many other storytelling approaches exist, such as animation, simulation, and more flexible interactive systems. Additionally, our work deals specifically with narrative data. Future research can explore the effects of other storytelling forms across different types of stories. 

%% file: body/7_Conclusion.tex
\section{Conclusion}
\label{sec:Conclusion}
In this work, we explored approaches for presenting abstract storyline visualization in VR environments. We discussed our design considerations and demonstrated a prototype of 3DStoryline visualization. We further evaluated its usability and discussed our findings and insights through this research. What we want to share at the end of this study is that immersive storytelling holds great potential for sharing and communicating insights. When creating such systems, designers should be very clear about the data type, the message of the stories, and the story pieces that can be used to support the message. These elements shape the story and determine the environment that would best host such stories.